\documentclass[journal,twoside,web]{ieeecolor}
\usepackage{tmi}
\usepackage{cite}
\usepackage{amsmath,amssymb,amsfonts}
\usepackage{algorithmic}
\usepackage{graphicx}
\usepackage{textcomp}

\def\BibTeX{{\rm B\kern-.05em{\sc i\kern-.025em b}\kern-.08em
T\kern-.1667em\lower.7ex\hbox{E}\kern-.125emX}}
\usepackage{hyperref}
\usepackage{booktabs}
\usepackage{float}
\usepackage{subfigure}
\usepackage{array}

\def\ab#1{{#1}}

\newcommand{\plans}{p\in\mathcal{P}}
\newcommand{\bb}{\mathbf{b}}

\newcommand{\ObjMatrix}{\mathbf{C}}
\newcommand{\x}{\mathbf{x}}
\newcommand{\xhat}{\hat{\x}}
\newcommand{\optObj}{\ObjMatrix\x}
\newcommand{\optObjHat}{\ObjMatrix\xhat}
\newcommand{\A}{\mathbf{A}}
\newcommand{\bfb}{\mathbf{b}}

\newcommand{\one}{\mathbf{e}}
\newcommand{\bfAlph}{\boldsymbol{\alpha}}
\newcommand{\alphaStar}{\bfAlph^*}
\newcommand{\bfp}{\mathbf{p}}
\newcommand{\bPrimep}{\bb'\bfp}
\newcommand{\maxSig}{\zeta}
\newcommand{\bfSig}{\boldsymbol{\sigma}}

\newcommand{\lowGap}{\boldsymbol{\delta}}
\newcommand{\smallNum}{\epsilon\one}

\newcommand{\pat}{p}
\newcommand{\roiOpt}{r}
\newcommand{\roiOpts}{\mathcal{R}_p}

\newcommand{\oarOpts}{\mathcal{I}_p}
\newcommand{\targetOpt}{t}
\newcommand{\targetOpts}{\mathcal{T}_p}
\newcommand{\targetOptPenalty}{\theta_\targetOpt}

\newcommand{\optStructs}{\mathcal{O}_p}

\newcommand{\voxel}{v}
\newcommand{\voxels}{\mathcal{V}}

\newcommand{\optDose}{d_\voxel}
\newcommand{\w}{w_b}
\newcommand{\beamlet}{b}
\newcommand{\beamlets}{\mathcal{B}}

\newcommand{\DOpt}[1]{\text{D}_\text{#1}}  
\newcommand{\Dmean}{$\DOpt{mean}$}  
\newcommand{\DOptcc}{$\DOpt{0.1cc}$}  
\newcommand{\DOptOne}{$\DOpt{1}$}  
\newcommand{\DOptNineFive}{$\DOpt{95}$}  
\newcommand{\DOptNineNine}{$\DOpt{99}$}  

\newcommand{\fixedThreshold}{f}

\newcommand{\obj}{g_m}
\newcommand{\objPred}{\hat{g}_m}
\newcommand{\objPair}{\left(\obj, \objPred\right)}
\newcommand{\setOfObjectiveFunctions}{m\in\mathcal{M}_\pat}
\newcommand{\DOptiffFunction}{\obj - \objPred}
\newcommand{\relativeDiffFunction}{\frac{\obj - \objPred}{\objPred}}
\newcommand{\optFun}[2]{\underset{\setOfObjectiveFunctions}{#1}\left(#2\right)}

\newcommand{\roiOptVoxels}{\voxel\in\voxels_\roiOpt}

\newcommand{\targetOptInTargets}{\targetOpt\in\targetOpts}

\newcommand{\roiOptInRois}{\roiOpt\in\roiOpts}

\newcommand{\undersetMax}[2]{\underset{#1}{\text{max}}\left\{#2\right\}}
\newcommand{\undersetMean}[2]{\underset{#1}{\text{mean}}\left\{#2\right\}}

\newcommand{\oneSide}[1]{\text{max}\left\{0,\ #1\right\}}

\newcommand{\GmaxConAbs}{\optObjHat + \maxSig\one}
\newcommand{\GmeanConAbs}{\optObjHat + \bfSig + \lowGap}
\newcommand{\GmaxConRel}{\optObjHat\odot(\one + \maxSig\one ) }
\newcommand{\GmeanConRel}{\optObjHat\odot(\one + \bfSig + \lowGap )}

\newcommand{\GdualAbs}{\bfAlph'\optObjHat-\bPrimep}

\newcommand{\GmaxNormAbs}{\bfAlph'\one = 1}
\newcommand{\GmeanNormAbs}{\bfAlph \le \one}
\newcommand{\GmaxNormRel}{\bfAlph'\optObjHat = 1}
\newcommand{\GmeanNormRel}{\bfAlph\odot\optObjHat \le \one}
\newcommand{\alphaLB}{\bfAlph \ge \smallNum}
\newcommand{\alphaCXLB}{\bfAlph\odot\optObjHat \ge \smallNum}

\newcommand{\genMeanForward}[1]{
	$\begin{aligned}
		 &  \underset{\mathbf{x}, \bfSig, \lowGap}{\text{min}}
		 & & \one'\bfSig + \smallNum'\lowGap\\
		 & \text{s.t.}
		 & &  \optObj = {#1} \\ 
		 & & & \A\x = \bfb  \\ 
		 & & & \x \ge \bf{0}\\
		 & & &  \bfSig \ge \bf{0} \\
		 & & &  \lowGap \le \bf{0} \\
	\end{aligned}$
}

\newcommand{\genMaxForward}[1]{
	$\begin{aligned}
		 &  \underset{\mathbf{x}, \maxSig}{\text{min}}
		 & & \maxSig\\
		 & \text{s.t.}
		 & &  \optObj \le {#1}\\
		 & & & \A\x = \bfb\\ 
		 & & & \x \ge \bf{0}\\
		 &\\
		 &
	\end{aligned}$
}

\newcommand{\genDual}[4]{
	$\begin{aligned}
		 & \underset{\boldsymbol{\alpha},\mathbf{p}}{\text{min}}
		 & &  {#1} \\
		 & \text{s.t.}
		 & &  \mathbf{C}' \boldsymbol{\alpha} \ge  \mathbf{A'p}\\ 
		 & & &  {#2}  \\
		 & & & {#4} \\
	\end{aligned}$
}

\newcommand{\genDualMax}[3]{
	$\begin{aligned}
		 & \underset{\boldsymbol{\alpha},\mathbf{p}}{\text{min}}
		 & &  {#1} \\
		 & \text{s.t.}
		 & &  \mathbf{C}' \boldsymbol{\alpha} \ge  \mathbf{A'p} \\
		 & & &  {#2}  \\
		 & & & \boldsymbol{\alpha} \ge \mathbf{0}
	\end{aligned}$
}

\begin{document}
	\title{OpenKBP-Opt: An international and reproducible evaluation of 76 knowledge-based planning pipelines}
	\author{
Aaron Babier, Rafid Mahmood, Binghao  Zhang, Victor G. L. Alves, Ana Maria Barragán-Montero, Joel Beaudry, Carlos E. Cardenas, Yankui Chang, Zijie Chen, Jaehee Chun, Kelly Diaz, Harold David Eraso, Erik Faustmann, Sibaji Gaj, Skylar Gay, Mary Gronberg, Bingqi Guo, Junjun He, Gerd Heilemann, Sanchit Hira, Yuliang Huang, Fuxin Ji, Dashan Jiang, Jean Carlo Jimenez Giraldo, Hoyeon Lee, Jun Lian, Shuolin Liu, Keng-Chi Liu, José Marrugo, Kentaro Miki, Kunio  Nakamura, Tucker Netherton, Dan Nguyen, Hamidreza Nourzadeh, Alexander F. I. Osman, Zhao Peng, José Darío Quinto Muñoz, Christian Ramsl, Dong Joo Rhee, Juan David Rodriguez, Hongming Shan, Jeffrey V. Siebers, Mumtaz H. Soomro, Kay Sun, Andrés Usuga Hoyos, Carlos Valderrama, Rob Verbeek, Enpei Wang, Siri Willems, Qi Wu, Xuanang Xu, Sen Yang, Lulin Yuan, Simeng Zhu, Lukas Zimmermann, Kevin L. Moore, Thomas G.  Purdie, Andrea L. McNiven, Timothy C. Y. Chan
	\thanks{Due to space constraints, funding information and author affiliations for this work appear in the acknowledgement section.}}

	\maketitle

	\begin{abstract}
		We establish an open framework for developing plan optimization models for knowledge-based planning (KBP) in radiotherapy. Our framework includes reference plans for 100 patients with head-and-neck cancer and high-quality dose predictions from 19 KBP models that were developed by different research groups during the OpenKBP Grand Challenge. The dose predictions were input to four optimization models to form 76 unique KBP pipelines that generated 7600 plans. The predictions and plans were compared to the reference plans via: dose score, which is the average mean absolute voxel-by-voxel difference in dose a model achieved; the deviation in dose-volume histogram (DVH) criterion; and the frequency of clinical planning criteria satisfaction. We also performed a theoretical investigation to justify our dose mimicking models. The range in rank order correlation of the dose score between predictions and their KBP pipelines was 0.50 to 0.62, which indicates that the quality of the predictions is generally positively correlated with the quality of the plans. Additionally, compared to the input predictions, the KBP-generated plans performed significantly better (P$<$0.05; one-sided Wilcoxon test) on 18 of 23 DVH criteria. Similarly, each optimization model generated plans that satisfied a higher percentage of criteria than the reference plans. Lastly, our theoretical investigation demonstrated that the dose mimicking models generated plans that are also optimal for a conventional planning model. This was the largest international effort to date for evaluating the combination of KBP prediction and optimization models. In the interest of reproducibility, our data and code is freely available at \href{https://github.com/ababier/open-kbp-opt}{https://github.com/ababier/open-kbp-opt}.
	\end{abstract}



	\section{Introduction}\label{sec:introduction}
\begin{figure}[htb]
		\centering
		\includegraphics[width=\linewidth]{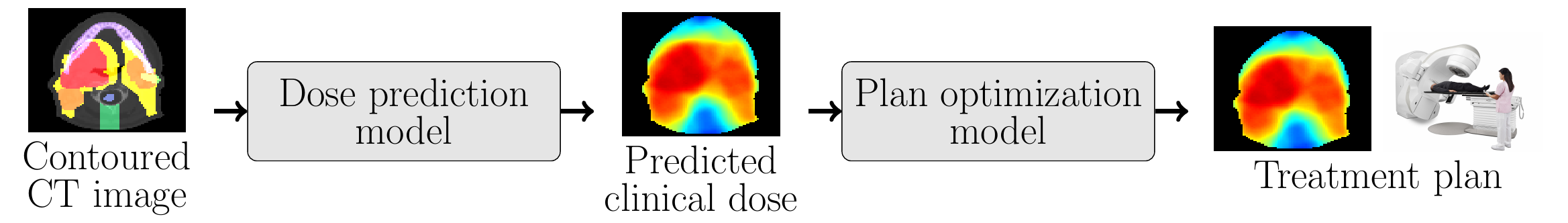}
		\caption{Overview of a complete knowledge-based planning pipeline.}
		\label{fig:pipelines}
	\end{figure}

	Automated radiotherapy planning is transforming clinical practice and personalized cancer treatment~\cite{Moore:2019aa}. The most common type of automated planning is knowledge-based planning, which leverages knowledge derived from historical clinical treatment plans to generate new treatment plans without human intervention~\cite{Cornell:2020aa,Kaderka:2021aa,McIntosh:2021aa}. Most common KBP methods can be thought of as a two-stage pipeline that first predicts the dose that should be delivered to a patient, and then converts that prediction into a treatment plan via optimization (Figure~\ref{fig:pipelines}). Both stages of this pipeline, which are active areas of research, can significantly affect the quality of generated treatment plans~\cite{Babier:2020ab}. The contributions of this paper are twofold: 1) to provide data that supports KBP optimization research at scale and 2) to establish a connection between dose mimicking (a type of KBP optimization) and conventional planning methods. We expand on the impact of these contributions throughout this paper.

	Comparing the quality of competing KBP models from the research community is difficult because the vast majority of research is conducted with large private datasets, as noted in several reviews~\cite{Hussein:2018aa,Ge:2019aa,Wang:2020aa,Momin:2021aa}. To help address this issue, the Open Knowledge-Based Planning (OpenKBP) Grand Challenge was organized to facilitate the largest international effort to date for developing and comparing dose prediction models on a single open dataset~\cite{Babier:2021aa}. \ab{The OpenKBP dataset, which includes data for 340 head-and-neck patients undergoing intensity modulated radiotherapy (IMRT), is limited to dose prediction research (i.e., it is incompatible with KBP optimization research). Although there are still no open datasets for KBP optimization research, there are two open datasets that support research in other areas of plan optimization}~\cite{Craft:2014aa,Breedveld:2017aa}. However, it is challenging to use these datasets in KBP plan optimization research for two reasons. First, neither dataset includes dose predictions, which are the input to KBP plan optimization models. Second, they are smaller (\ab{123 patients across both datasets}), span multiple sites (\ab{prostate, liver, head-and-neck}), and multiple modalities (\ab{CyberKnife, volumetric modulated arc therapy,  proton therapy, IMRT}). While such \ab{a diversity in cases} is important to demonstrate the robustness and generalizability of optimization algorithms across sites and modalities, this same diversity is a disadvantage when it comes to training dose prediction models, since there is insufficient data for any one site-modality pair~\cite{sSize}.


	Most KBP pipelines are developed as fully-automated pipelines that can replace human treatment planners in the planning process~\cite{McIntosh:2017aa,Fan:2018aa,Bai:2020aa,Wortel:2021}. These approaches have demonstrated promising results in \ab{prospective research studies} where a sizeable portion of KBP-generated plans were considered inferior to human-generated plans, \ab{which suggests that there is an opportunity for improvement}~\cite{Cornell:2020aa,McIntosh:2021aa}. In those cases, making manual adjustments to the KBP-generated plan is non-trivial because they are generated by \emph{fully-automated} pipelines that rely on the quality of the data. In contrast to fully automated pipelines, \emph{semi-automated} pipelines rely on both the quality of data and human expertise, which puts less reliance on the data. For example, a semi-automated KBP pipeline could enable human planners to improve upon a KBP-generated plan via an intuitive process (e.g., inverse planning) and thereby provide a pipeline that leverages human expertise, models, and data. In the KBP literature, however, there are relatively few papers that describe tools that humans can intuitively interact with in semi-automated KBP pipeline~\cite{Li:2017aa,Babier:2018a,Bohara:2020aa,Zhang:2021}.

	In this paper, we extend the results from the OpenKBP Grand Challenge, which we call OpenKBP, with an international validation of 76 KBP pipelines. We made this extension, which we call OpenKBP-Opt, open to provide a benchmark for KBP optimization research and to lower the barriers for contributing to this research area. We also demonstrate how KBP plan optimization models can be used to initialize the conventional planning process (i.e., inverse planning) with good patient-specific parameters (i.e., objective weights) \ab{and provide the means for} a semi-automated KBP pipeline. Identifying this relationship provides a mechanism for transforming existing KBP optimization models, which are generally fully-automated pipelines that impede manual intervention, into semi-automated pipelines that promote human planners to improve upon a KBP-generated plan via inverse planning (i.e., a familiar and intuitive process). The data and code to reproduce this paper is publicly available at \href{https://github.com/ababier/open-kbp-opt}{https://github.com/ababier/open-kbp-opt}.


	\section{Methods and materials}\label{sec:methods-and-materials}
\begin{figure}[htb]
		\centering
		\includegraphics[width=\linewidth]{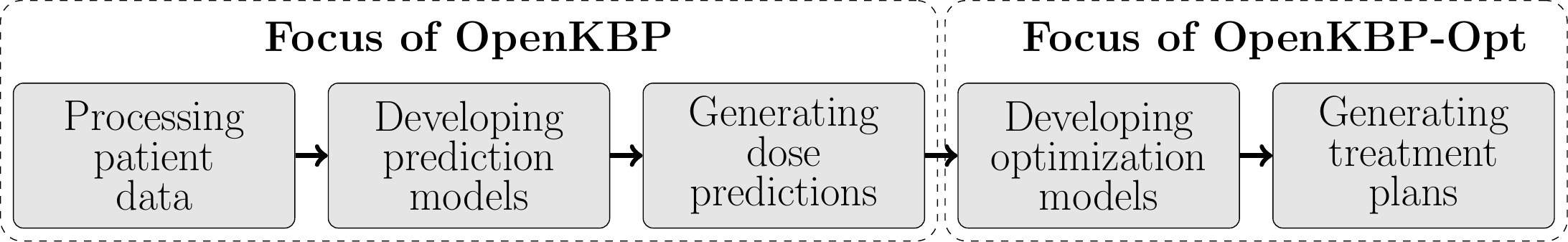}
		\caption{An overview of our methods. A full description of each component is provided in the corresponding subsection.}
		\label{fig:methodsOverview}
	\end{figure}

	Figure~\ref{fig:methodsOverview} summarizes the overall methodological approach into five components. The first three components (i.e., processing data, developing dose prediction models, and generating KBP dose predictions) are based on the results of the OpenKBP Grand Challenge. The final two components (i.e., developing plan optimization models and generating KBP treatment plans) are an extension of the OpenKBP Grand Challenge and the focus of this paper. Below, we describe all five components and our analysis.

	\subsection{Processing patient data}\label{subsec:process-data}
	We obtained data for 340 patients ($n=340$) with head-and-neck cancer from the OpenKBP Grand Challenge. The data consisted of a training set ($n=200$), a validation set ($n=40$), and a testing set ($n=100$). The plans were delivered via 6 MV step-and-shoot IMRT from nine equidistant coplanar beams at angles 0$^\circ$, 40$^\circ$, \ldots, 320$^\circ$. Those beams were divided into a set of beamlets $\mathcal{B}$, which make up a fluence map. The relationship between the intensity $w_b$ of beamlet $b$ and dose $d_v$ deposited to voxel $v$ was determined using the influence matrix $D_{v,b}$ generated by the \textsf{IMRTP} library from \texttt{A\ Computational\ Environment\ for\ Radiotherapy\ Research}~\cite{CERR} using \texttt{MATLAB}, and it is given by

	\begin{equation}
		\label{IM}
		d_v = \sum\limits_{b \in \mathcal{B}} D_{v,b}w_b.
	\end{equation}

	\subsection{Developing dose prediction models}\label{subsec:develop-dose-prediction-methods}
	All dose prediction models used in this paper were developed in the OpenKBP Grand Challenge~\cite{Babier:2021aa}. During the challenge, teams developed dose prediction models using identical training and validation datasets \ab{with access only to ground truth data (i.e., dose) for the training set}. Every dose prediction model used a neural network architecture that was based on either a U-Net~\cite{Unet}, V-Net~\cite{Vnet}, or Pix2Pix~\cite{isola:2017image} architecture. Many of the best performing models also used other generalizable techniques like ensembles~\cite{Nguyen:2021uv}, one-cycle learning~\cite{Zimmermann:2021aa}, radiotherapy-specific loss functions~\cite{Gronberg:2021aa}, and deep supervision~\cite{Liu:2021aa}.

    All teams competed to develop models that minimize one of two pre-defined error metrics that quantified the difference between the reference dose and a KBP-generated dose (i.e., KBP prediction or plan dose). The metrics were: 1) dose error, which was the mean absolute voxel-by-voxel difference between two dose distributions, and 2) dose-volume histogram (DVH) error, which was the absolute difference between a DVH point from two dose distributions. The DVH error was evaluated on two and three DVH points for each organ-at-risk (OAR) and target, respectively. The OAR DVH points were \ab{the \Dmean\ and \DOptcc, which was the mean dose delivered to the OAR and the maximum dose delivered to 0.1cc of the OAR, respectively}. The target DVH points were the \DOptOne, \DOptNineFive\ab{,} and \DOptNineNine, which was the dose delivered to 1\% (99$^\textrm{th}$ percentile)\ab{,} 95\% (5$^\textrm{th}$ percentile)\ab{,} and 99\% (1$^\textrm{st}$ percentile) of voxels in the target, respectively. The models were ranked according to: 1) dose score, which was the average dose error of a model, and 2) DVH score, which was the average DVH error of a model.

	\subsection{Generating KBP dose predictions}\label{subsec:predict-test-set-dose-distributions}
	In this paper, the OpenKBP organizers collaborated with teams that competed in the OpenKBP Grand Challenge. \ab{The 28 teams that completed the ﬁnal phase of the OpenKBP Grand Challenge were invited to participate in the OpenKBP-Opt project, and 21 of those teams agreed to participate.} We obtained the dose predictions from all teams for each patient in the test set to create a set of 2100 dose predictions (21 different predictions for each of the 100 patients). We observed that two models produced dose scores that were over two standard deviations (6.3 Gy) above the mean (4.0 Gy), whereas the rest were within half a standard deviation (1.6 Gy) of the mean. Thus, we omitted those two outlier models and proceeded with only 19 KBP models ($n=1900$ predictions).

	\subsection{Developing plan optimization models}\label{subsec:develop-plan-optimization-methods}
	Next, we formulated four dose mimicking models, which are a type of KBP optimization model. Each model used the same set of structures and objective functions that we described in Section~\ref{subsubsec:structures} and Section~\ref{subsubsec:objectives}, respectively. However, they differ in how they mimic (i.e., penalize differences) a specific dose distribution. In particular, they each have a different cost function, outlined in Section~\ref{subsubsec:model-formulations}. \ab{Note that in this paper the terms ``objective function'' and ``cost function'' refer to distinct concepts, and the cost functions in this paper are functions of objective functions.}


	\subsubsection{Structures}\label{subsubsec:structures} All of our optimization models used the same set of regions-of-interest (ROIs) $\roiOpts$ for each patient $\plans$ in our test set. The set $\roiOpts$ contains OARs $\oarOpts$, targets $\targetOpts$, and optimization structures $\optStructs$. The OARs contained in $\oarOpts$ were the brainstem, spinal cord, right parotid, left parotid, larynx, esophagus, and mandible. Each target $\targetOptInTargets$ was a planning target volume (PTV) with a dose level $\targetOptPenalty$, and those targets were the PTV56, PTV63, and PTV70. The optimization structures contained in $\optStructs$ were the limPostNeck, which was used to limit dose to the posterior neck, and six PTV ring structures (a $3~\text{mm}$ ring and a $6~\text{mm}$ ring for each target). These were the same structures used to generate the plans in the original OpenKBP dataset~\cite{Babier:2021aa}. Every ROI $\roiOptInRois$ was also divided into a set of voxels $\mathcal{V}_r$.

	\subsubsection{Objective functions}\label{subsubsec:objectives} Our models used the objective functions in Table~\ref{tab:doseObectives}. Each objective function quantified a different measure of the dose delivered to a single ROI $\roiOptInRois$ in a patient $\plans$, which we call an objective value. Specifically, the average and maximum objective values quantified the average dose and maximum dose delivered to an ROI $r$, respectively. The high and low conditional value at risk (CVaR) objective values quantified the average dose in ROI $r$ that was higher and lower, respectively than the dose threshold $\fixedThreshold$. 

	\begin{table}[htb]
		\centering
		\caption{The formulations for objective functions in our models.}
		\begin{tabular}{lcc}
			\toprule
			Name        & Objective function                                                    \\ \midrule
			Average dose   & $\undersetMean{\roiOptVoxels}{\optDose}$                              \\
			Maximum dose   & $ \undersetMax{\roiOptVoxels}{\optDose} $                             \\
			High CVaR dose & $ \undersetMean{\roiOptVoxels}{\oneSide{\optDose - \fixedThreshold}}$ \\
			Low CVaR dose  & $ \undersetMean{\roiOptVoxels}{\oneSide{\fixedThreshold - \optDose}}$ \\\bottomrule
		\end{tabular}
		\label{tab:doseObectives}
	\end{table}

	In total, we considered 107 objectives functions: seven per OAR, three per target, and seven per optimization structure. The objective functions for each OAR were the mean dose; maximum dose; and high CVaR dose with thresholds $\fixedThreshold$ equal to 0.25, 0.50, 0.75, 0.90, and 0.975 of the maximum predicted dose to that structure. The objective functions for each target were the maximum dose, low CVaR dose with a threshold equal to the dose level of the target (i.e.,  $\fixedThreshold=\targetOptPenalty$), and a high CVaR dose with a threshold $\fixedThreshold$ equal to 1.05 of the dose level of the target (i.e., $\fixedThreshold=1.05\targetOptPenalty$). The objective functions for each optimization structure were the same as the OAR objective functions. Not all patients had all ROIs, so the models associated with those patients had fewer than 107 objective functions.

	\subsubsection{Model formulations}\label{subsubsec:model-formulations} Our KBP optimization models performed dose mimicking to generate plans with optimized objective values that closely matched the input objective values from a dose prediction. To streamline our model formulation, let each $\setOfObjectiveFunctions$ denote one of the 107 objective functions (as outlined in Section~\ref{subsubsec:objectives}). Let  $\obj$ and $\objPred$ be objective values of their corresponding objective functions evaluated over the optimized plan and predicted dose, respectively. In all models, the cost functions were formulated such that lower values of $\obj$ were favored over higher values.

	Table~\ref{tab:models} presents the cost functions of our dose mimicking models. Each model minimized either the mean or max difference between all corresponding pairs $\objPair$ of the objective values, which were quantified via an absolute (e.g., $\DOptiffFunction$) or relative (e.g., $(\obj - \objPred)/\objPred$) difference measure, resulting in four dose mimicking models. In the mean difference models, we chose to prioritize the positive differences (i.e., where the optimized plan objective value was higher than the predicted dose objective value) more than the negative differences, which we assigned a small positive weight $\epsilon$ ($\epsilon = 0.0001$ in our experiments). This was done to incentivize the model to do at least as well as the dose prediction before striving to outperform the dose prediction on other objective functions. In contrast, the max difference models used only a single term because the max difference naturally incentivizes the model to outperform the prediction only once the plan outperforms the prediction across all objective values (i.e., when $\obj \le \objPred, \forall \setOfObjectiveFunctions$).

	\begin{table}[htb]
		\centering
		\caption{The cost functions for each model that minimize mean absolute (MeanAbs), max absolute (MaxAbs), mean relative (MeanRel), and max relative (MaxRel) differences between the optimized and predicted objective values $\objPair$.}
		\begin{tabular}{l c}
			\toprule
			& Optimization model cost function                                                                            \\  \midrule\addlinespace[1ex]
			MeanAbs & $\optFun{\text{mean}}{\DOptiffFunction}^+\ +\ \epsilon\ \optFun{\text{mean}}{\DOptiffFunction}^-$           \\\addlinespace[1ex]
			MaxAbs  & $\optFun{\max}{\DOptiffFunction}$                                                                           \\\addlinespace[1ex]

			MeanRel & $\optFun{\text{mean}}{\relativeDiffFunction}^+\ +\ \epsilon\ \optFun{\text{mean}}{\relativeDiffFunction}^-$ \\\addlinespace[1ex]

			MaxRel  & $\optFun{\max}{\relativeDiffFunction}$                                                                      \\\bottomrule
		\end{tabular}
		\label{tab:models}
	\end{table}

	The main constraint in all four models was a constraint to limit plan complexity. In particular, the sum-of-positive gradients (SPG)~\cite{Craft:2007aa} of all plans generated by the models was constrained to be less than or equal to 65, which was a constraint in the reference plans~\cite{Babier:2021aa}. The remaining constraints were simply auxiliary constraints (including auxiliary variables) used to linearize both the objective and cost functions (i.e., the formulations in Table~\ref{tab:doseObectives} and Table~\ref{tab:models}). The optimization models were all formulated in \textsf{Python 3.7} using \textsf{OR-Tools 8.2} and solved using \textsf{Gurobi 9.1} (Gurobi Optimization, TX, US) on a single computer with an Intel i7-8700K (6-Core $3.7$ GHz) CPU and $16$ GB of random access memory. Default parameters were used with the \textsf{Gurobi} solver except for \emph{Crossover} set to $0$, \emph{Method} set to $2$, and \emph{BarConvTol} set to $0.0001$, which were selected based on past experience to improve solve time without compromising solution quality.

	\subsection{Generating KBP treatment plans}\label{subsec:optimize-test-set-treatment-plans}
\begin{figure}[htb]
		\centering
		\includegraphics[width=\linewidth]{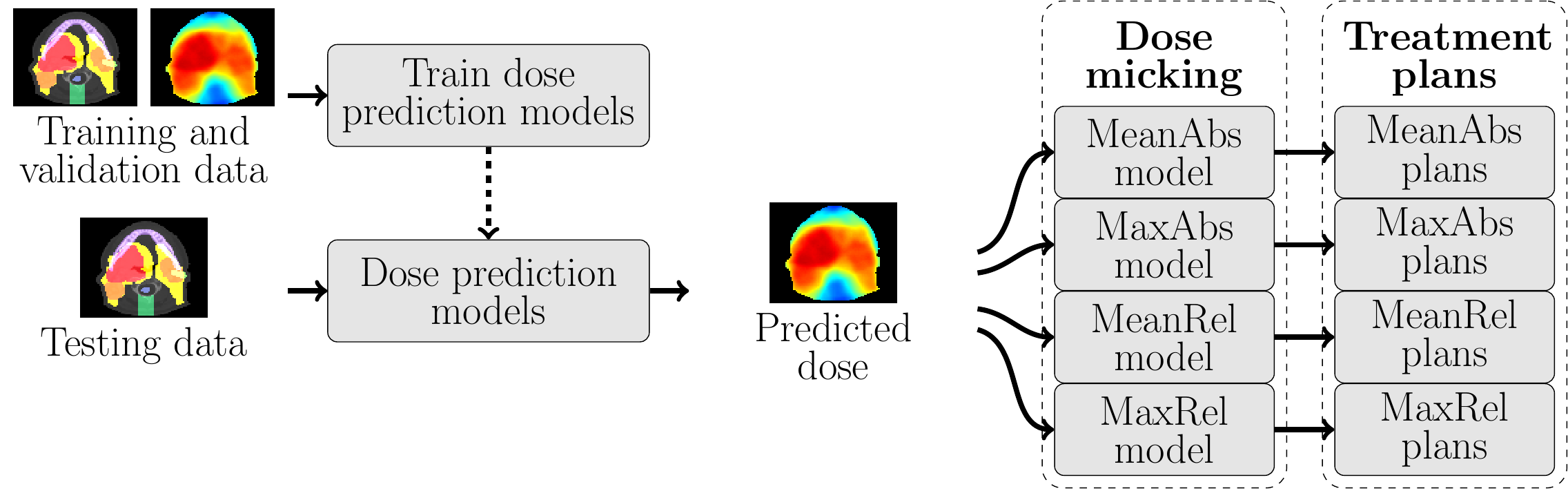}
		\caption{An overview of our process. First, dose prediction models were trained on out-of-sample data. Those models were used to predict dose for input to dose mimicking optimization models to generate KBP plans.}
		\label{fig:methods}
	\end{figure}

	Next, we assembled 76 KBP pipelines by combining the 19 dose prediction models with each of the four dose mimicking models. Each pipeline was applied to the 100 patients in the testing set, resulting in 7600 KBP plans (see Figure~\ref{fig:methods}). We used these plans in our analysis to measure the quality of the respective KBP models. We refer to the four plans generated from each dose prediction as the MeanAbs, MaxAbs, MeanRel, and MaxRel plans.

	\ab{Altogether, after completing the process in Figure~\ref{fig:methods}, we had dose distributions for a set of reference plans ($n=100$), predictions ($n=1900$), and KBP plans generated by four dose mimicking models ($n = 4 \times 1900$). The reference plans are the plans that were released as part of the OpenKBP Grand Challenge, and the predictions are dose distributions that were submitted by 19 teams in the final testing phase of OpenKBP. In general, there will be differences between the reference plan, prediction, and KBP plan dose distributions. Differences between a dose prediction and its corresponding KBP plan are due to factors including prediction noise and deliverability of the dose prediction. Differences between a KBP plan and its corresponding reference plan reflect different trade-offs in the cost function used to generate these plans.}

	\subsection{Analysis}\label{subsec:analysis}
	We conducted three analyses to measure model performance in terms of dose error, DVH point differences, and clinical criteria satisfaction. We also investigated the theoretical connection between our dose mimicking models and inverse planning. Finally, we summarized empirical optimization metadata.

	\subsubsection{Dose score and error} We evaluated the KBP models using the dose score and dose error as defined in Section~\ref{subsec:develop-dose-prediction-methods}. We calculated the Spearman rank order correlation of the dose score between the prediction models and corresponding KBP pipelines. The distribution of dose error was visualized using a box plot. A one-sided Wilcoxon signed-rank test was used to determine whether the dose error of the optimization models was the same (null hypothesis) or lower (alternative hypothesis)  than the dose predictions models. For all hypothesis tests in this paper, $P < 0.05$ was considered significant.

	\subsubsection{DVH point differences} To measure the relative quality of dose distributions from a clinical perspective, we examined the distribution of DVH point differences between the reference and KBP-generated dose. The differences were evaluated over the DVH points listed in Section~\ref{subsec:develop-dose-prediction-methods} and visualized using boxplots. We used the one-sided Wilcoxon signed-rank test to determine whether the dose generated by all optimization models performed the same (null hypothesis) or better (alternative hypothesis) than the dose predictions. This test was chosen to evaluate the aggregate performance of all optimization models relative to the predictions. Lower values were better for $\DOpt{mean}$, $\DOpt{0.1cc}$, and $\DOpt{1}$; higher values were better for $\DOpt{95}$ and $\DOpt{99}$.

	\subsubsection{Expected criteria satisfaction}\label{subsubsec:expected-criteria-satisfaction} As another measure of plan quality, we examined the proportion of clinical criteria that were satisfied by the reference plans and KBP-generated dose. One criterion was evaluated for each ROI (see Table~\ref{tab:criteria}). We tabulated the proportion of criteria that were satisfied by the reference plans, dose predictions, MeanAbs plans, MaxAbs plans, MeanRel plans, MaxRel plans, and the plans from the KBP pipeline that satisfied the most clinical criteria overall. We also plotted the proportion of OAR, target, and all ROI clinical criteria that each of the 76 KBP pipelines achieved.

	\begin{table}[htb]
		\centering
		\caption{The clinical criteria that we used to evaluate dose distributions. \ab{Before evaluating these criteria, we reinstated any overlap between targets that was removed.}}
		\begin{tabular}{llc}
			\toprule
			\multicolumn{2}{l}{Structures} & Criteria \\ \midrule
			\multicolumn{2}{l}{OARs} & \multicolumn{1}{l}{} \\
			& Brainstem     & $\DOpt{0.1cc} \le 50.0 \text{ Gy}$ \\
			& Spinal cord   & $\DOpt{0.1cc} \le 45.0 \text{ Gy}$ \\
			& Right parotid & $\DOpt{mean} \le 26.0 \text{ Gy}$  \\
			& Left parotid  & $\DOpt{mean} \le 26.0 \text{ Gy}$  \\
			& Esophagus     & $\DOpt{mean} \le 45.0 \text{ Gy}$  \\
			& Larynx        & $\DOpt{mean} \le 45.0 \text{ Gy}$  \\
			& Mandible      & $\DOpt{0.1cc} \le 73.5 \text{ Gy}$ \\
			\multicolumn{2}{l}{Targets} & \multicolumn{1}{l}{} \\
			& PTV56         & $\DOpt{99} \ge 53.2 \text{ Gy}$    \\
			& PTV63         & $\DOpt{99} \ge 59.9 \text{ Gy}$    \\
			& PTV70         & $\DOpt{99} \ge 66.5 \text{ Gy}$    \\\bottomrule
		\end{tabular}
		\label{tab:criteria}
	\end{table}

	\subsubsection{Theoretical analysis of dose mimicking models}\label{subsubsec:connection-to-inverse-planning}\ab{To justify our choice of dose mimicking models, we conducted a theoretical analysis into their structure using linear programming duality theory~\cite[Chapter~4]{ORIntro}. This analysis was based on previous literature that showed a connection between Benson’s method~\cite{Benson:1978wm}, which identifies efficient solutions to multi-objective optimization models, and estimating the weights for inverse planning~\cite{taewoo}. We were motivated to conduct a similar analysis as in Chan~\textit{et al.}~\cite{taewoo} because our dose mimicking models are similar to the formulations in Benson~\cite{Benson:1978wm}. In particular, we linearized the dose mimicking models, took their duals, and related the dual variables to objective function weights in a conventional multi-objective planning problem depicted in model~\eqref{eq:conPlanning}.}
	\begin{equation}
		\label{eq:conPlanning}
		\begin{aligned}
			&\underset{g}{\text{minimize}}
			&&\sum\limits_{\setOfObjectiveFunctions}\hat{\alpha}_m\obj, \\
			& \text{subject to}
			& & SPG \le 65,\\
			& & &  \text{Auxiliary constraints to linearize}\\\addlinespace[-0.5ex]
			& & & \text{functions in Table}~\ref{tab:doseObectives}\text{ and }\ref{tab:models}.
		\end{aligned}
	\end{equation}

	\subsubsection{Optimization metadata}\label{subsubsec:optimization-metadata} Lastly, we summarized the metadata that each optimization model generated. In particular, we evaluated the average proportion of objective weight that each model assigned to OAR, target, and optimization structure objective functions. Additionally, we recorded the average, first quartile, and third quartile solve time.


	\section{Results}\label{sec:results}
	In this section, we summarize the performance of the 19 dose predictions models, four dose mimicking models, and 76 KBP pipelines.

	\subsection{Dose score and error}\label{subsec:kbp-errors}
	Table~\ref{tab:prediction-plan-changes} summarizes the rank order correlation between the dose prediction models and their corresponding KBP pipelines. We found that the rank of a prediction model is positively correlated with its corresponding KBP pipeline rank. However, there was a wide range in correlation from 0.50 to 0.62. This demonstrates that high quality predictions are correlated with high quality plans, but this result also indicates that a prediction model that outperforms a competitor will not always generate better plans. Additionally, the KBP plans generated by an optimization model that evaluated relative differences (i.e., MeanRel and MaxRel) achieved higher rank order correlations than their counterparts that evaluated absolute differences (i.e., MeanAbs and MaxAbs).

	\begin{table}[htb]
		\centering
		\caption{Each dose mimicking model is compared to the predictions in terms of median rank change and rank order correlation.}
		\setlength{\tabcolsep}{.175em}
		\begin{tabular}{lcccc}
			\toprule
			& MeanAbs & MaxAbs & MeanRel & MaxRel \\\midrule
			Rank order correlation & 0.53    & 0.50   & 0.62    & 0.59   \\
			Rank order $P$-value   & 0.019   & 0.030  & $  0.005$ & 0.008  \\\bottomrule
		\end{tabular}
		\label{tab:prediction-plan-changes}
	\end{table}

\begin{figure}[htb]
		\centering
		\includegraphics[width=1\linewidth]{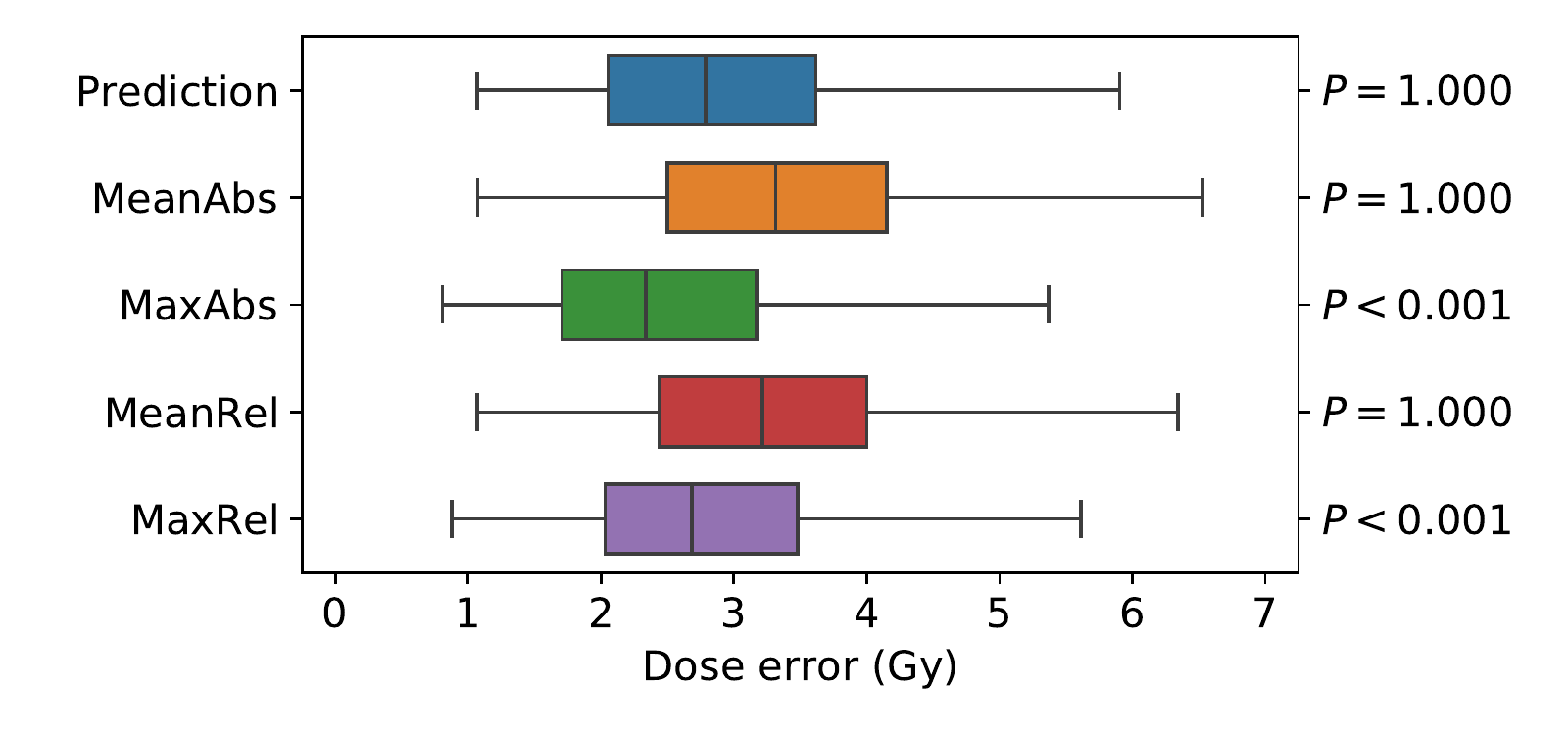}
		\caption{The distribution of dose error over all KBP-generated dose \ab{($n=1900$ points in each box)}. Boxes indicate median and interquartile range (IQR). Whiskers extend to the minimum of 1.5 times the IQR and the most extreme outlier.}
		\label{fig:kbp-error}
	\end{figure}

	The dose errors of predictions and KBP plans are shown in Figure~\ref{fig:kbp-error}. Two of the four sets of KBP plans had a median dose error that was lower than the median dose error of the predictions (2.79 Gy), implying that it is possible for optimization models to generate dose distributions that more closely resemble the reference plan dose, compared to dose predictions. These two models also achieved a significantly lower error ($P\le0.001$) than predictions. The MaxAbs model achieved the lowest median dose error (2.34 Gy).


	\subsection{DVH point differences}\label{subsec:dvh-point-differences}
\begin{figure*}[htb]
		\centering
		\subfigure[OAR $\DOpt{mean}$]{
			\includegraphics[width=0.4675\linewidth]{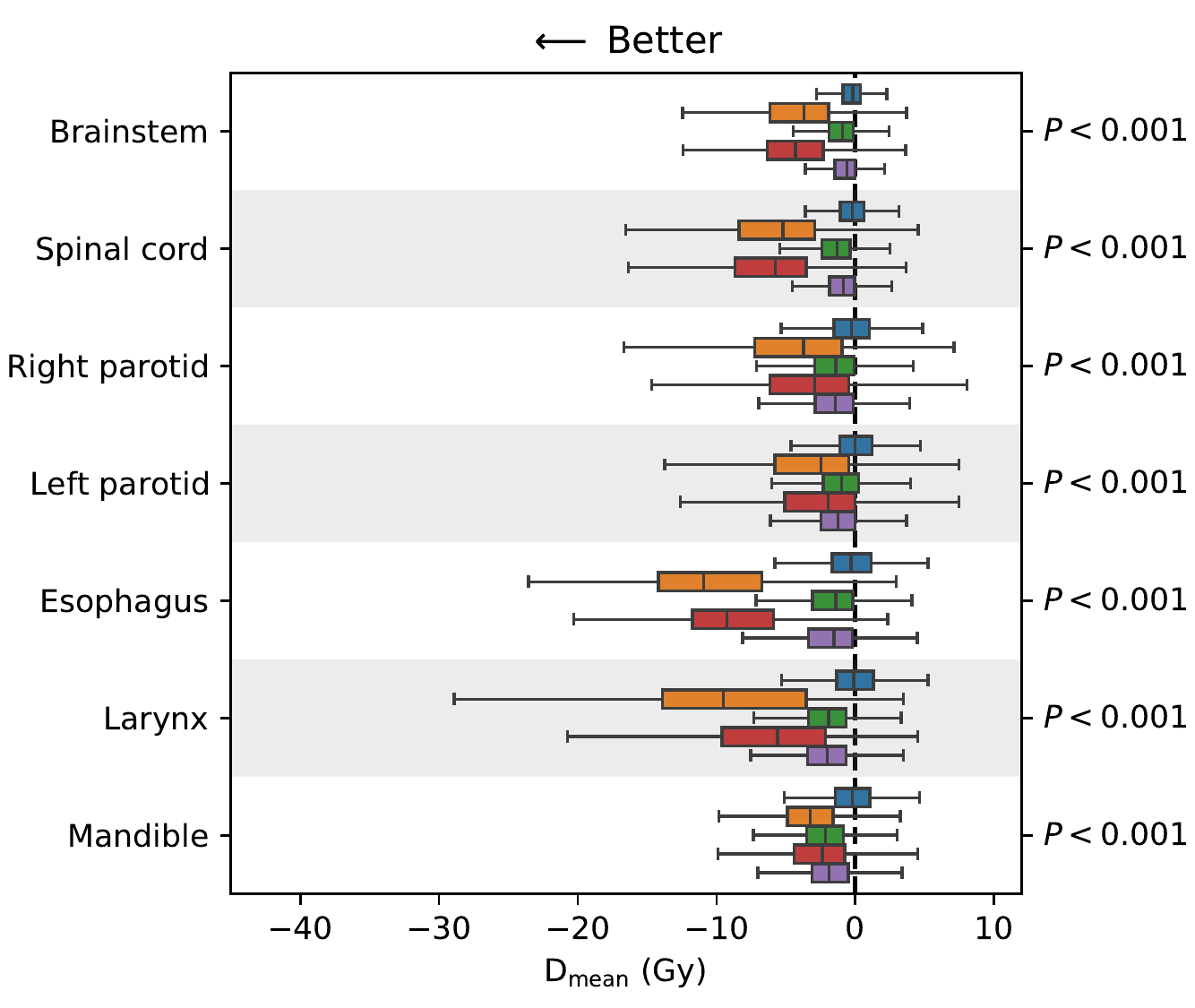}}
		\subfigure[OAR $\DOpt{0.1cc}$]{
			\includegraphics[width=0.4675\linewidth]{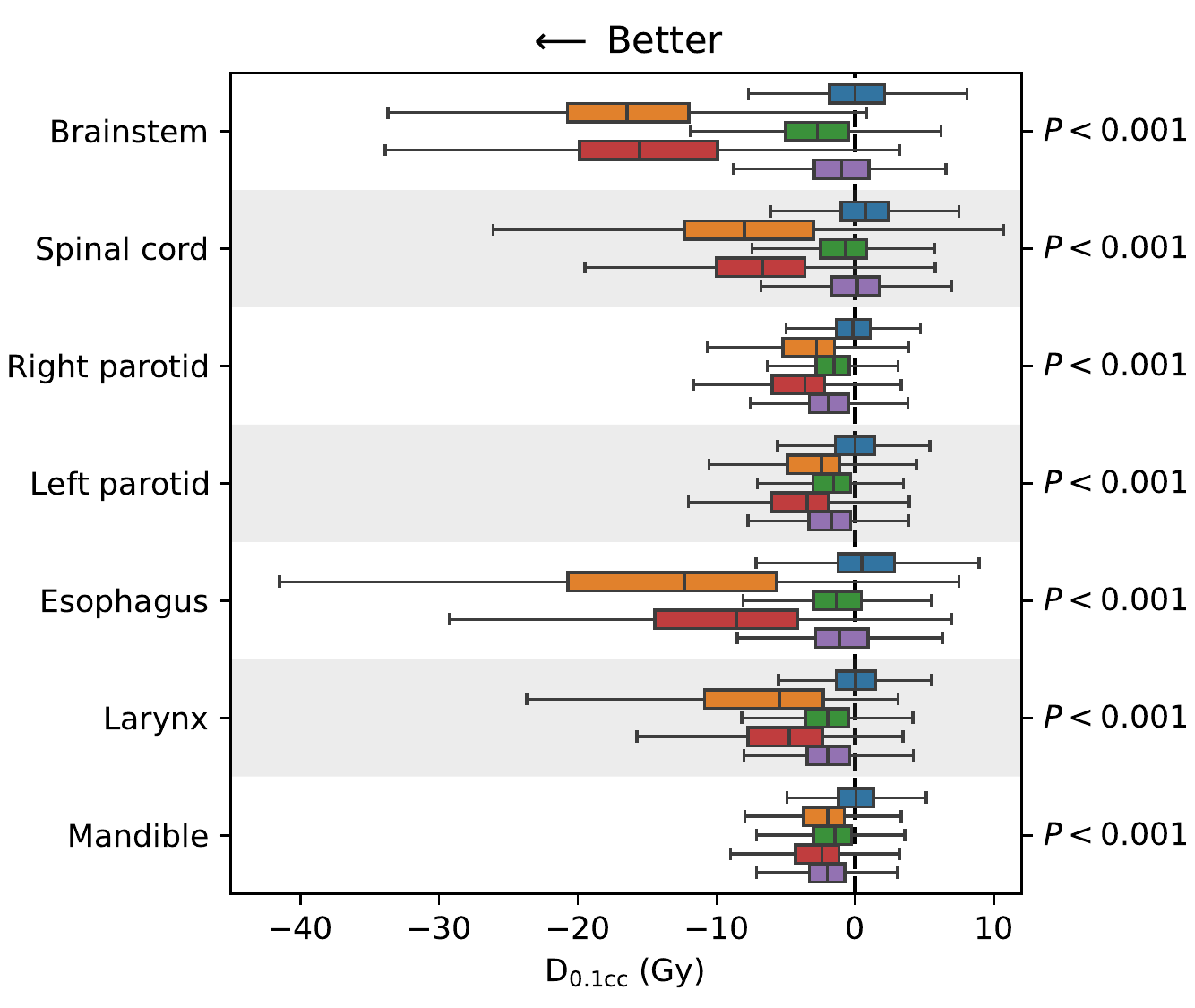}}\\
		\subfigure[Target $\DOpt{1}$]{
			\includegraphics[width=0.31\linewidth]{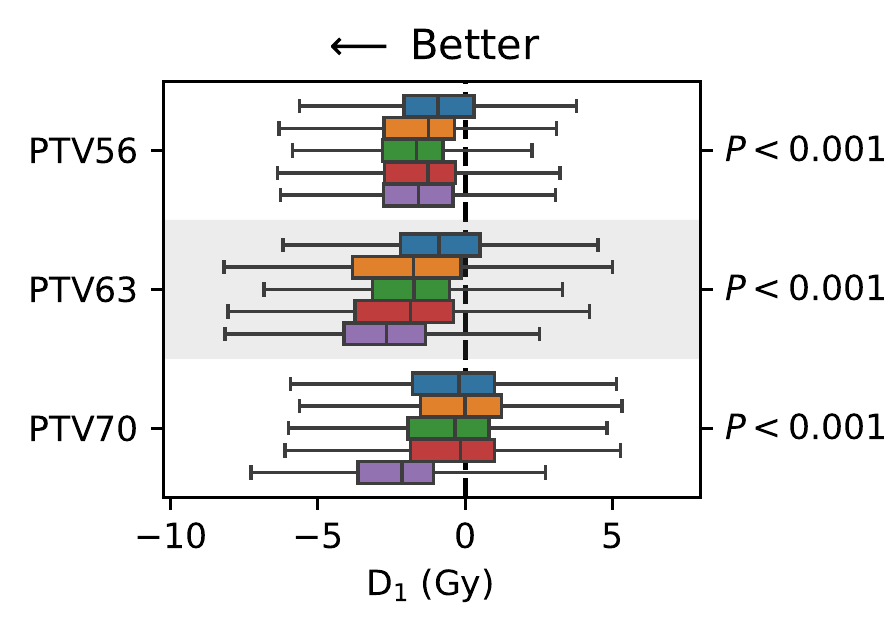}}
		\subfigure[Target $\DOpt{95}$]{
			\includegraphics[width=0.31\linewidth]{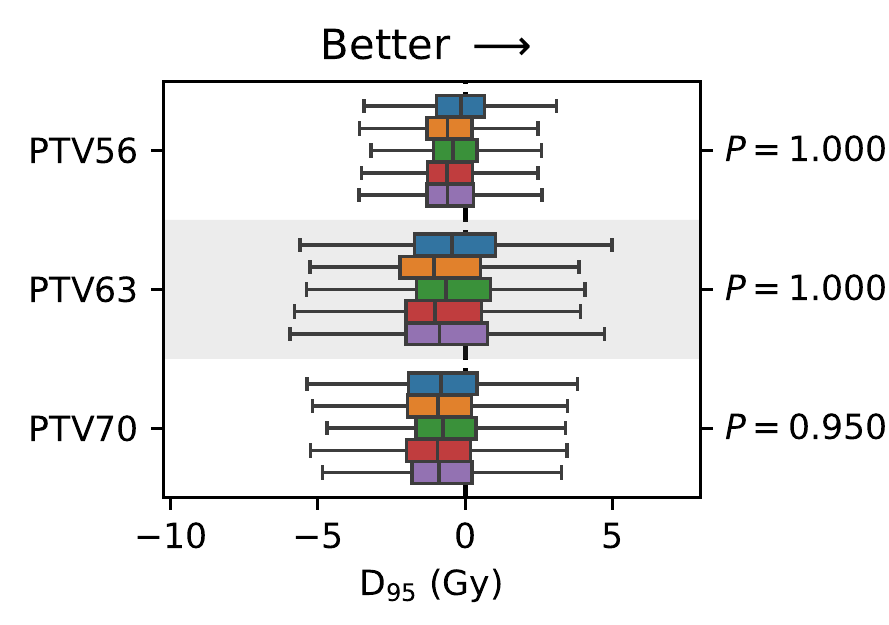}}
		\subfigure[Target $\DOpt{99}$]{
			\includegraphics[width=0.31\linewidth]{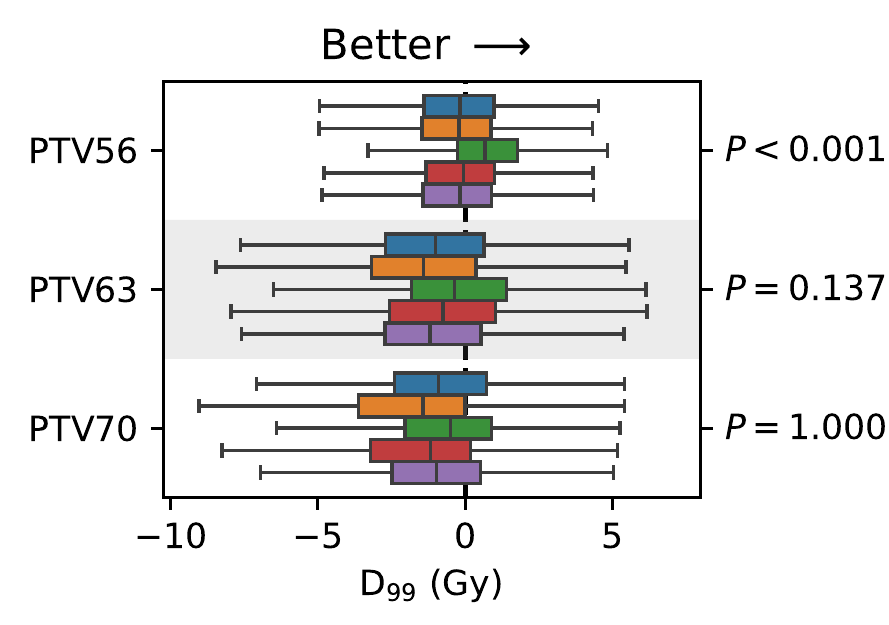}}\\
		\includegraphics[width = 0.65\linewidth]{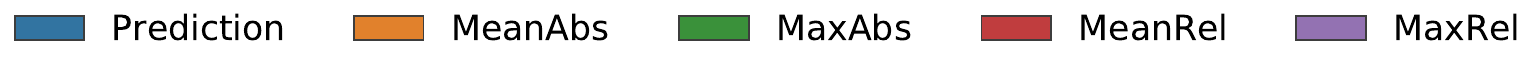}
		\caption{The distribution of DVH point differences between the reference dose and each set of KBP-generated dose. Negative differences indicate cases where the KBP-generated dose had a lower DVH points than the reference dose. Boxes indicate median and IQR. Whiskers extend to the minimum of 1.5 times the IQR and the most extreme outlier.}
		\label{fig:plan-quality}
	\end{figure*}

	Figure~\ref{fig:plan-quality} shows the DVH point differences between the reference dose and either the predicted dose or KBP plan dose. In general, dose mimicking tends to produce a plan dose that is significantly better than the dose it received as input from a dose prediction model. In particular, the KBP plan dose is significantly better on 18 of the 23 DVH points than the predicted dose (all OAR points and four target points). The five DVH points where the plans were not significantly better are the three $\DOpt{95}$ points and two $\DOpt{99}$ points.

	\subsection{Expected criteria satisfaction}\label{subsec:expected-criteria-satisfaction2}
	In Table~\ref{tab:criteria-satisfaction}, we compare the percentage of criteria that were satisfied by the reference plans ($n=100$), the predictions ($n=1900$), the plans generated by each of the four dose mimicking models ($n = 4 \times 1900$), and the plans generated by the top performing KBP pipeline ($n=100$). We use the term baselines to refer to the reference dose and dose predictions collectively. The top performing KBP pipeline (denoted ``Best'' in Table~\ref{tab:criteria-satisfaction}) was defined as the single pipeline \ab{(i.e., the combination of one dose prediction model and one dose mimicking model)} whose plans satisfied the most clinical criteria. Of all dose mimicking models, the MaxRel and MeanAbs models generated plans that satisfied the fewest (69.8\%) and most (72.9\%) ROI clinical criteria, respectively. For comparison, predictions only satisfied 66.2\% of all clinical criteria, which was 3.5 percentage points lower than the reference plans (69.7\%). The best KBP pipeline, which used the MeanAbs model and one of the 19 prediction models (discussed later), satisfied 77.0\% of all ROI clinical criteria.

	In general, clinical criteria satisfaction varied across each ROI criterion. The brainstem, spinal cord, esophagus, and mandible criteria were each satisfied more than 85\% of the time across all the baselines and our dose mimicking models in Table~\ref{tab:criteria-satisfaction}. The right parotid, left parotid, and larynx were satisfied less than 40\% of the time for the two baselines. In contrast, each of our four KBP models generated a higher average criteria satisfaction for these ROIs compared to the baselines. In fact, some were substantially higher. For example, the average criteria satisfaction of the MeanAbs model on the larynx was 71.5\%, compared to an average of 36.2\% for the baselines. In aggregate over all 19 prediction models, the performance of the four dose mimicking model was comparable or slightly worse than the reference dose in terms of criteria satisfaction in the targets. However, the best KBP pipeline outperformed the baselines on all criteria.

	\begin{table*}[htb]
		\centering
		\caption{The percentage of clinical criteria satisfied in each set of KBP-generated dose. Note that ``Best'' is defined as the top performing KBP pipeline that generated plans that satisfied the most ROI clinical criteria. The highest percentage of satisfied criteria is bolded in each row.}
		\begin{tabular}{lllccccccccc}
			\toprule
			& & & \multicolumn{2}{c}{Baselines} & & \multicolumn{4}{c}{Dose mimicking models}
			&
			&
			\\ \cline{4-5} \cline{7-10}\addlinespace[1ex]
			\multicolumn{2}{l}{} & & Reference & Prediction & & MeanAbs & MaxAbs & MeanRel & MaxRel & & Best \\\midrule
			\multicolumn{2}{l}{OARs} & & & & & & & & & & \\
			& Brainstem     & & 96.6 & 97.3 & & \textbf{100.0} & 99.5 & \textbf{100.0} & 98.5 & & \textbf{100.0} \\
			& Spinal cord   & & 95.5 & 92.7 & & 99.7           & 97.3 & \textbf{100.0} & 95.6 & & \textbf{100.0} \\
			& Right parotid & & 32.3 & 32.7 & & \textbf{46.1}  & 38.9 & 45.0           & 38.0 & & 41.4           \\
			& Left parotid  & & 30.6 & 30.1 & & \textbf{43.7}  & 35.0 & 41.9           & 35.0 & & 40.8           \\
			& Esophagus     & & 93.0 & 92.7 & & \textbf{100.0} & 95.2 & \textbf{100.0} & 97.3 & & \textbf{100.0} \\
			& Larynx        & & 37.7 & 34.7 & & \textbf{71.5}  & 44.9 & 58.8           & 44.6 & & 67.9           \\
			& Mandible      & & 87.5 & 89.4 & & \textbf{99.6}  & 98.7 & 99.2           & 99.0 & & 93.1           \\
			\multicolumn{2}{l}{Targets} & & & & & & & & & & \\
			& PTV56         & & 91.2 & 85.8 & & 83.3           & 91.8 & 84.1           & 84.6 & & \textbf{96.7}  \\
			& PTV63         & & 90.5 & 86.2 & & 82.2           & 89.6 & 84.8           & 84.8 & & \textbf{92.9}  \\
			& PTV70         & & 64.0 & 45.7 & & 37.2           & 51.6 & 40.1           & 47.7 & & \textbf{66.0}  \\
			\multicolumn{2}{l}{All} & & & & & & & & & & \\
			& OARs          & & 65.5 & 65.1 & & \textbf{77.1}  & 70.6 & 75.3           & 70.2 & & 74.5           \\
			& Targets       & & 79.4 & 68.7 & & 63.3           & 74.2 & 65.3           & 68.8 & & \textbf{82.8}  \\
			& ROIs          & & 69.7 & 66.2 & & 72.9           & 71.7 & 72.3           & 69.8 & & \textbf{77.0}  \\\bottomrule
		\end{tabular}
		\label{tab:criteria-satisfaction}
	\end{table*}

\begin{figure}[H]
		\centering
		\subfigure[All OAR criteria]{
			\includegraphics[width=0.95\linewidth]{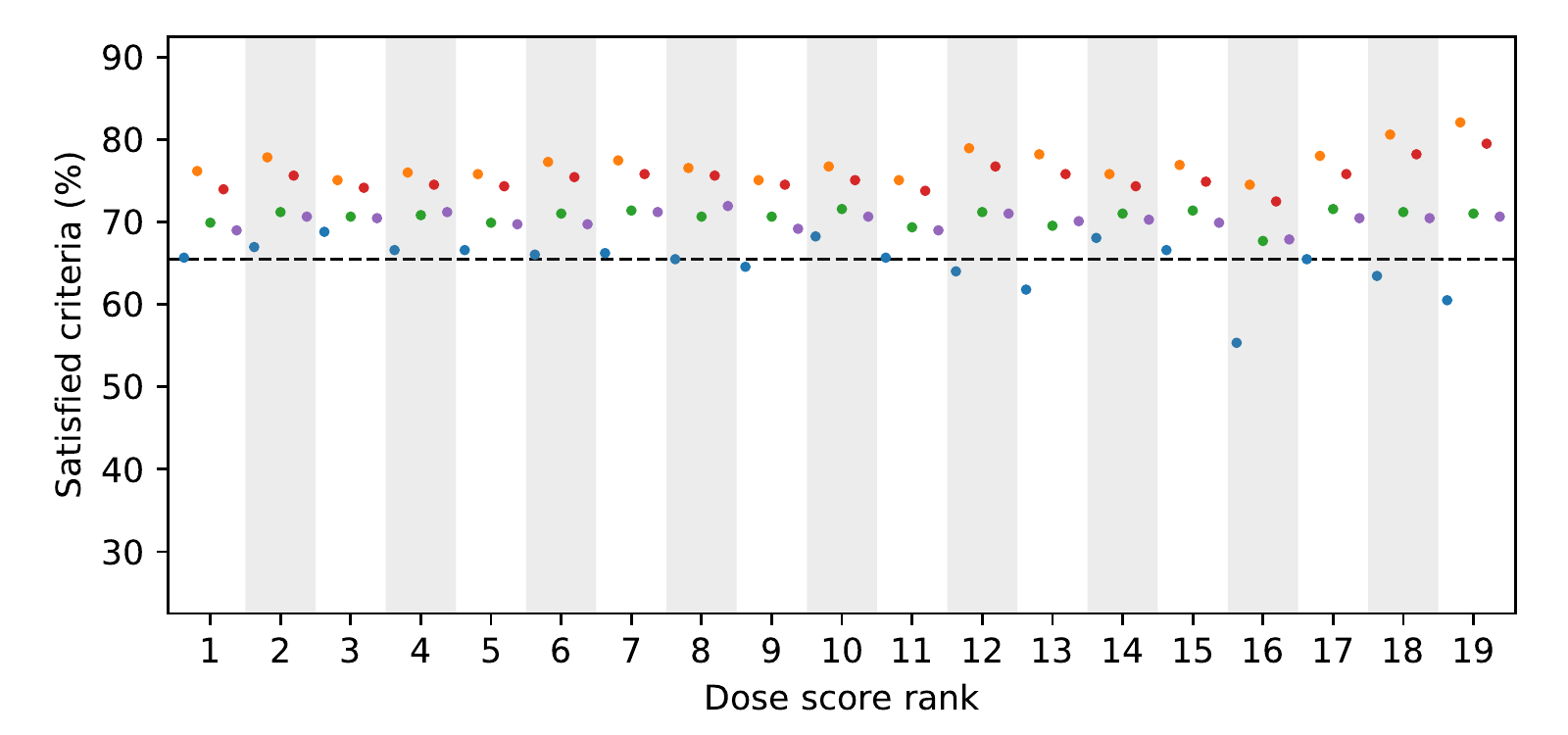}
			\label{fig:oar-criteria}}
		\subfigure[All target criteria]{
			\includegraphics[width=0.95\linewidth]{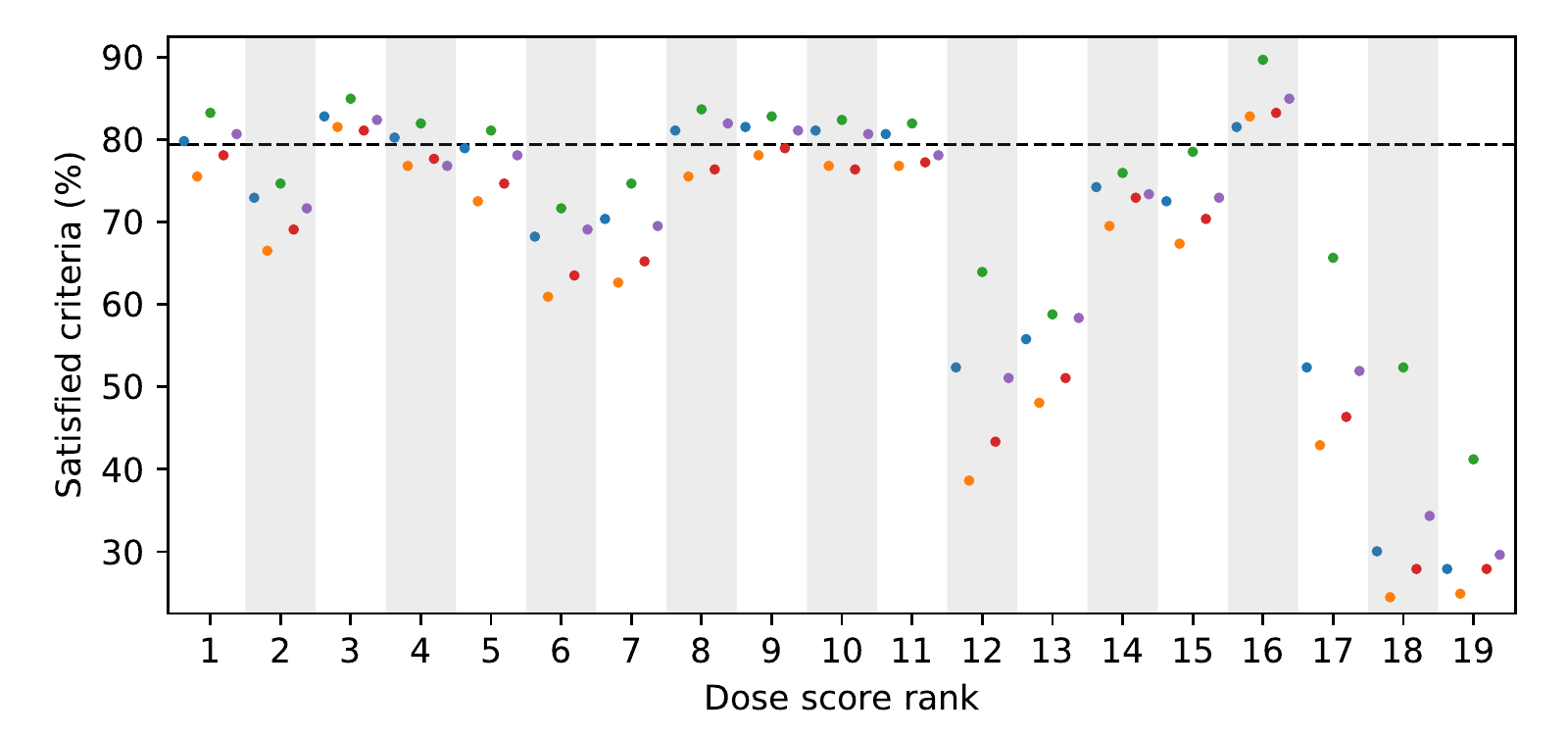}
			\label{fig:target-criteria}}
		\subfigure[All ROI criteria]{
			\includegraphics[width=0.95\linewidth]{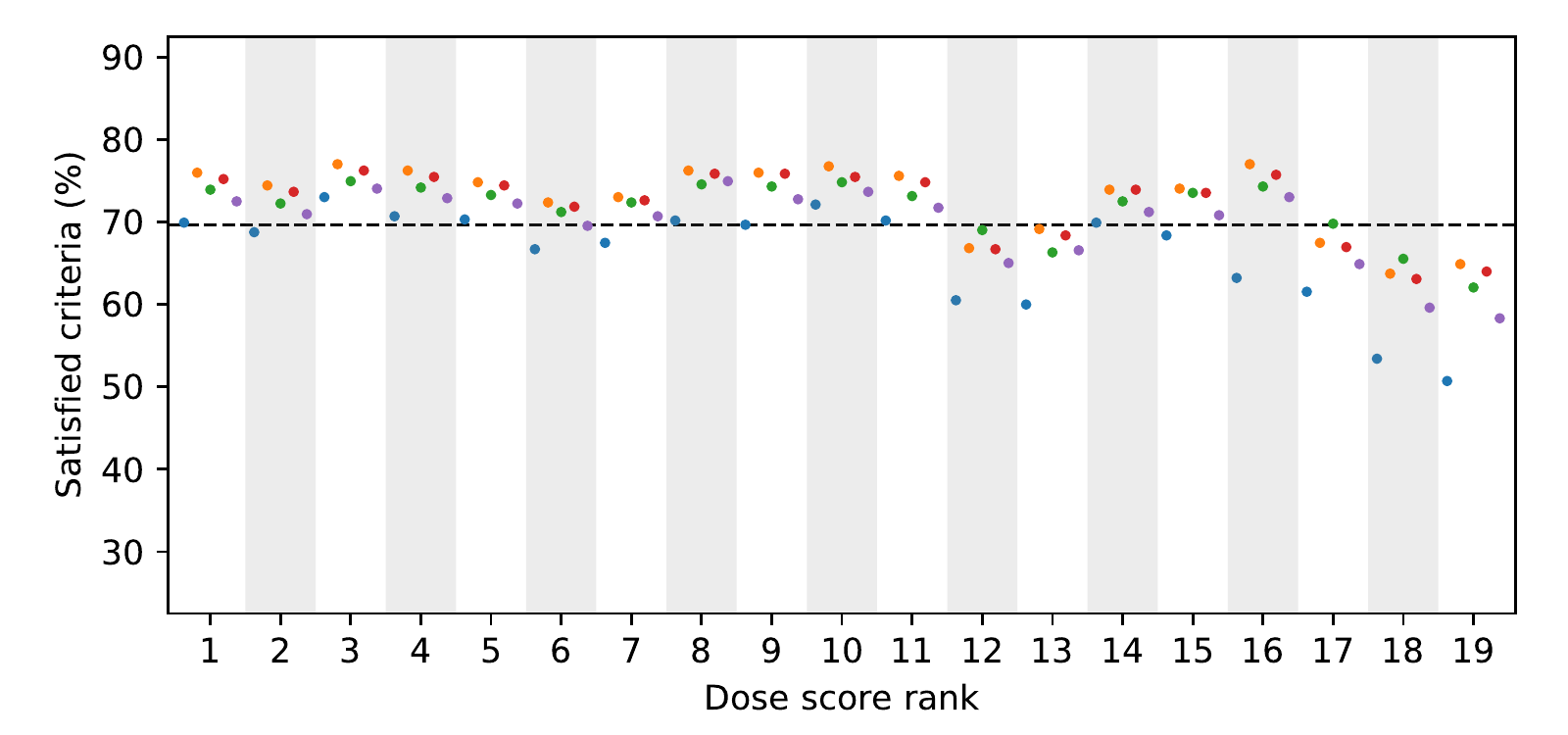}
			\label{fig:all-criteria2}}
		\includegraphics[width=1\linewidth]{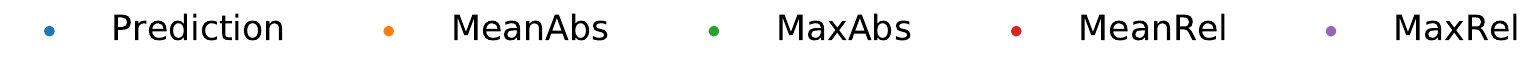}
		\caption{The percentage of (a) OAR, (b) Target, and (c) all ROI clinical criteria that were satisfied by each KBP pipeline. The points indicate the percentage of satisfied criteria \ab{for $n=100$ patients}. A dashed line indicates the percentage of criteria satisfied by reference plans.}
		\label{fig:plan-criteria}
	\end{figure}

	Figure~\ref{fig:plan-criteria} summarizes the clinical criteria that were satisfied by each of the 76 KBP pipelines that we evaluated. The MeanAbs model generated plans that satisfied more criteria than the other three optimization models for 16 of the 19 dose prediction models (see Figure~\ref{fig:all-criteria2}). Additionally, the pipelines that used better prediction models (i.e., dose score rank closer to 1) generally produced plans with higher criteria satisfaction. Interestingly, the best performing KBP pipeline (the last column of Table~\ref{tab:criteria-satisfaction}) used the dose prediction model that ranked 16$^\textrm{th}$ in terms of dose score. The \ab{spread} in OAR criteria satisfaction across all 19 models (55.4\% to 82.1\%) was lower than that of target criteria satisfaction (24.5\% to 89.7\%), see Figure~\ref{fig:oar-criteria} and Figure~\ref{fig:target-criteria}, respectively. Note that the poor performing KBP pipelines used the 12$^\textrm{th}$, 13$^\textrm{th}$, 17$^\textrm{th}$, 18$^\textrm{th}$, and 19$^\textrm{th}$ ranked dose prediction models. Since the columns in Table~\ref{tab:criteria-satisfaction} included all KBP pipelines, these poor performing models contributed to low performance on the target criteria. In contrast, many of the KBP pipelines that used the top ranked models prediction models clearly performed much better on target criteria.

	\subsection{Theoretical analysis of dose mimicking models}

	The inverse planning model~\eqref{eq:conPlanning} is shown in model~\eqref{eq:conPlanningMatrix} in vector and matrix notation following Chan~\textit{et al.}~\cite{taewoo}.
	\begin{equation}
		\label{eq:conPlanningMatrix}
		\begin{aligned}
			& \underset{\mathbf{x}}{\text{minimize}}
			& & \hat{\bfAlph}'\optObj \\
			& \text{subject to}
			& & \A\x = \bfb,\\
			& & & \x \ge \mathbf{0}.
		\end{aligned}
	\end{equation}
	The objective functions are the rows of matrix $\ObjMatrix$ and the objective function weights are represented by the vector $\hat{\bfAlph}$. The decision variables, which include the fluence variables ($\w\ \forall\beamlet\in\beamlets$) and auxiliary variables are represented by vector $\x$. The SPG and auxiliary constraints are encoded in the matrix $\A$ and vector $\bfb$.

	Table~\ref{tab:matrixModels} presents the formulations of the four dose mimicking models and their respective dual models. \ab{The positive and negative differences between optimized objective values $\optObj$ and predicted objective values $\optObjHat$ are represented by vectors $\bfSig$ and $\lowGap$, respectively. The max difference between the optimized and predicted objective values is expressed as a scalar $\maxSig$. The dual variables of the dose mimicking models are denoted  $\bfAlph$ and $\bfp$. The vectors of all $0$ and $1$ are denoted by $\mathbf{0}$ and $\one$, respectively. The symbol $\odot$ denotes element-wise multiplication of two vectors and prime denotes the transpose operator.}

	\ab{Next, we complete our theoretical analysis}. By Proposition 5 from~\cite{taewoo}, it follows that an optimal decision vector $\x^*$ from each dose mimicking model is also optimal for the inverse planning model~\eqref{eq:conPlanningMatrix}) with an optimal dual vector $\alphaStar$ as objective weights (i.e., $\x^*$ is an optimal solution for model~\eqref{eq:conPlanningMatrix} when $\hat{\bfAlph} = \alphaStar$). This result means that the solution to each dose mimicking model is also optimal to the inverse planning model with a particular set of objective function weights.

%

	\begin{table*}[htb]
		\centering
		\caption{The dose mimicking models used in this paper are presented in matrix notation with their corresponding dual models.}
		\begin{tabular}{m{1.25cm}llll}
			\toprule
			& MeanAbs & MaxAbs & MeanRel & MaxRel \\\midrule
			Dose mimicking model & \genMeanForward{\GmeanConAbs} & \genMaxForward{\GmaxConAbs}
			& \genMeanForward{\GmeanConRel}
			& \genMaxForward{\GmaxConRel}
			\\ \midrule
			Dual model & \genDual{\GdualAbs}{\GmeanNormAbs}{\bfSig}{\alphaLB} & \genDualMax{\GdualAbs}{\GmaxNormAbs}{\maxSig}
			&
			\genDual{\GdualAbs}{\GmeanNormRel}{\bfSig}{\alphaCXLB}
			& \genDualMax{\GdualAbs}{\GmaxNormRel}{\maxSig}
			\\\bottomrule
		\end{tabular}
		\label{tab:matrixModels}
	\end{table*}

	\subsection{Optimization metadata}\label{subsec:kbp-plan-metadata}
	In Table~\ref{tab:metadata}, we present metadata that was generated by each optimization model, which assigned a different proportion of weight to the objectives for each group of ROIs (i.e., OARs, targets, optimization structures). The models that evaluate relative differences (i.e., MeanRel and MaxRel) spread the proportion of weight relatively evenly between the OAR and target objectives, but the other two models assigned the majority of the weight to target objectives with no more than 0.018 weight to OARs. Additionally, the optimization structures generally received the smallest proportion of weight with the exception of the MaxAbs model, which assigned more weight to optimization structure objectives (0.170) than OAR objectives (0.011). There is also a wide range in average solve time of the models (222 seconds to 393 seconds). On average, the MaxAbs model was the fastest.

	\begin{table}[htb]
		\centering
		\caption{A summary of the metadata that each optimization model generated after optimizing 1900 plans.}
		\setlength{\tabcolsep}{0.35em}
		\begin{tabular}{lcccc}
			\toprule
			& MeanAbs & MaxAbs & MeanRel & MaxRel \\ \midrule
			\multicolumn{5}{l}{Objective weight} \\
			\quad OARs           & 0.018   & 0.011  & 0.554   & 0.417  \\
			\quad Targets        & 0.976   & 0.819  & 0.418   & 0.569  \\
			\quad Optimization   & 0.006   & 0.170  & 0.028   & 0.014  \\
			\multicolumn{5}{l}{Solve time (s)} \\
			\quad Average        & 389     & 222    & 367     & 393    \\
			\quad First quartile & 192     & 107    & 183     & 188    \\
			\quad Third quartile & 502     & 261    & 481     & 507    \\
			\bottomrule
		\end{tabular}
		\label{tab:metadata}
	\end{table}


	\section{Discussion}\label{sec:discussion}
	Knowledge-based planning research is flourishing. However, optimization models for KBP (e.g., dose mimicking) have received much less attention in the literature than dose prediction models. In this paper, we developed four dose mimicking models and evaluated their performance with 19 different dose prediction models, which were inputs to the optimization models. We showed that both the dose prediction model and optimization model contributed to considerable variation in the quality of plans generated by the corresponding KBP pipeline. Additionally, we conducted a theoretical investigation to show that our dose mimicking models generate plans that are optimal for a multi-objective inverse planning model with particular weights.

	Our data and code is published at \href{https://github.com/ababier/open-kbp-opt}{https://github.com/ababier/open-kbp-opt}.
	to enable others to reproduce our results, which meets the gold standard in reproducibility~\cite{Heil:2021tw}. Our data includes the first open dataset of predictions and reference plans to accompany CT images. We hope that this effort produces a common resource and lowers the barriers for future KBP optimization research, given that researchers must currently acquire their own private datasets and develop in-house prediction models before they can start testing new KBP optimization models.

	Our open dataset contains the data for 100 patients who were treated with IMRT and a sample of high quality dose predictions for those same patients. The dataset was curated for the purpose of developing new fluence-based KBP optimization models that use ROI masks, dose influence matrices, and a dose prediction. The dose predictions were generated by 21 dose prediction models that were developed by an international group of researchers, which provided a diverse sample of realistic inputs for a KBP optimization model. Two of those prediction models (20$^\textrm{th}$ and 21$^\textrm{th}$ ranked model) were removed from our analysis because their dose scores were low, \ab{which we elaborated on in Section~\ref{subsec:predict-test-set-dose-distributions}}. For completeness, however, those 200 predictions are also available as part of our dataset.

	We also performed a theoretical analysis to justify our dose mimicking models. Our key theoretical finding was that dose mimicking and conventional inverse planning are equivalent under certain specifications of the objective function weights. This allows us to interpret previous weight estimation techniques~\cite{taewoo} through the more intuitive lens of dose mimicking models. Finally, by connecting dose mimicking to inverse planning, there is the potential to convert fully-automated KBP pipelines into semi-automated pipelines. Specifically, we use dose mimicking to generate a high-quality plan with its corresponding objective weights, which can be used in an inverse planning model (i.e., model~\eqref{eq:conPlanningMatrix}). This is advantageous because it enables human planners to improve the quality of plans generated by KBP via a conventional inverse planning process. By enabling this intuitive human interaction, we create a semi-automated KBP pipeline that is aligned with a common belief that AI will augment, rather than replace, the duties of healthcare practitioners~\cite{Ahuja:2019aa}.

	Evaluating the performance of optimization models using many different dose predictions helps to identify interaction effects between these two stages of a KBP pipeline~\cite{Babier:2020ab}. For example, the 16$^\textrm{th}$ ranked model generated lower quality predictions (in terms of dose error) than most of its competitors. However, when used in a KBP pipeline with the right optimization model, in this case the MeanAbs model, it generated high quality plans that achieved more clinical criteria than any other KBP pipeline. In other words, the errors made by the 16$^\textrm{th}$ ranked model that contribute to its low prediction quality were corrected by the KBP optimization model. Since these interaction effects contribute to considerable variation in quality,  it is important to evaluate KBP optimization models across a diverse set of dose prediction models. Additionally, if we can understand what types of prediction error are most highly correlated with KBP plan quality we could propose better evaluation metrics to drive KBP prediction research towards making predictions that consistently translate into higher quality plans.

	As in the original OpenKBP challenge, a limitation of this work is that we use synthetic dose distributions (i.e., the reference dose) as a substitute for real clinical dose. Although these dose distributions were subject to less quality assurance than clinical plans, they were previously shown to be of similar quality~\cite{Babier:2021aa}. A second limitation of this work is that the dose prediction models were developed with the goal of optimizing the dose and DVH scores. There may be other scoring metrics that are better suited for developing a dose prediction model that excels in a KBP pipeline. This is a possible direction for future research. Lastly, this work only covers a single site and treatment modality. There is no guarantee that KBP optimization models that are developed with this dataset can generalize to other sites or treatment modalities.


	\section{Conclusion}\label{sec:conclusion}
	In this paper, we combined the dose predictions contributed by a large international team to several KBP optimization models, resulting in 76 KBP pipelines. This was the largest international effort to date on KBP pipeline evaluation. We found that the best performing pipeline significantly outperformed the baseline approaches. In the interest of reproducibility, our data and code is freely available.





	\section{Acknowledgments}\label{sec:acknowledgments}

	\footnotesize{

\thanks{Aaron Babier is with Department of Mechanical and Industrial Engineering, University of Toronto, Toronto, ON, Canada and Vector Institute, Toronto, ON, Canada (e-mail: \mbox{ababier@mie.utoronto.ca})}

\thanks{Rafid Mahmood and Binghao  Zhang are with Department of Mechanical and Industrial Engineering, University of Toronto, Toronto, ON, Canada (e-mail: \mbox{rafid.mahmood@mail.utoronto.ca};  \mbox{binghao.zhang@mail.utoronto.ca})}

\thanks{Victor G. L. Alves, Jeffrey V. Siebers,  and Mumtaz H. Soomro are with Department of Radiation Oncology, University of Virginia Health System, Charlottesville, VA, United States (e-mail: \mbox{VL8FN@hscmail.mcc.virginia.edu}; \mbox{JS2UB@hscmail.mcc.virginia.edu};  \mbox{ms2et@virginia.edu})}

\thanks{Ana Maria Barragán-Montero is with Department of Molecular Imaging Radiation Oncology, UCLouvain, Louvain-la-Neuve, Belgium (e-mail: \mbox{ana.barragan@uclouvain.be})}

\thanks{Joel Beaudry is with Department of Radiation Oncology, Memorial Sloan Kettering Cancer Center, New York, NY, United States (e-mail: \mbox{joelbeaudry@gmail.com})}

\thanks{Carlos E. Cardenas is with Department of Radiation Oncology, The University of Alabama at Birmingham, Birmingham, AL, United States (e-mail: \mbox{cecardenas@uabmc.edu})}

\thanks{Yankui Chang and Zhao Peng are with Department of Engineering and Applied Physics, University of Science and Technology of China, Hefei, China (e-mail: \mbox{cykanhui@mail.ustc.edu.cn};  \mbox{hnpz1232@mail.ustc.edu.cn})}

\thanks{Zijie Chen and Enpei Wang are with Shenying Medical Technology Co., Ltd., Shenzhen, Guangdong, China (e-mail: \mbox{chenzijie@sundymed.com};  \mbox{wangenpei172@126.com})}

\thanks{Jaehee Chun is with Department of Radiation Oncology, Yonsei University, Seoul, Korea (e-mail: \mbox{cjhsmile@gmail.com})}

\thanks{Kelly Diaz, Harold David Eraso, Jean Carlo Jimenez Giraldo, José Marrugo, José Darío Quinto Muñoz, Juan David Rodriguez, Andrés Usuga Hoyos,  and Carlos Valderrama are with Department of Physics , National University of Colombia, Medellín, Colombia (e-mail: \mbox{kmdiazd@unal.edu.co}; \mbox{hderasoe@unal.edu.co}; \mbox{jecjimenezgi@unal.edu.co}; \mbox{jlmarrugom@unal.edu.co}; \mbox{jdquintomu@unal.edu.co}; \mbox{judrodriguezar@unal.edu.co}; \mbox{yausugah@unal.edu.co};  \mbox{csvalderramap@unal.edu.co})}

\thanks{Erik Faustmann and Christian Ramsl are with Atominstitut, Vienna University of Technology, Vienna, Austria (e-mail: \mbox{erik.faustmann@gmail.com};  \mbox{christianramsl@gmail.com})}

\thanks{Sibaji Gaj, Bingqi Guo,  and Kunio  Nakamura are with Department of Biomedical Engineering, Cleveland Clinic, Cleveland, OH, United States (e-mail: \mbox{sibajigaj@gmail.com}; \mbox{guob@ccf.org};  \mbox{nakamuk@ccf.org})}

\thanks{Skylar Gay, Mary Gronberg, Tucker Netherton,  and Dong Joo Rhee are with Department of Radiation Physics, The University of Texas MD Anderson Cancer Center, Houston, TX, United States (e-mail: \mbox{skylar.sgay1@gmail.com}; \mbox{mpeters1@mdanderson.org}; \mbox{tnetherton@mdanderson.org};  \mbox{drhee1@mdanderson.org})}

\thanks{Junjun He is with Department of Biomedical Engineering, Shanghai Jiao Tong University, Shanghai, China (e-mail: \mbox{hejunjun@sjtu.edu.cn})}

\thanks{Gerd Heilemann is with Department of Radiation Oncology, Medical University of Vienna, Vienna, Austria (e-mail: \mbox{gerd.heilemann@meduniwien.ac.at})}

\thanks{Sanchit Hira is with Department of Biomedical Engineering, Johns Hopkins University, Baltimore, MD, United States (e-mail: \mbox{sanchithira24@gmail.com})}

\thanks{Yuliang Huang is with Department of Radiation Oncology, Peking University Cancer Hospital and Institute, Beijing, China (e-mail: \mbox{rrtwpw19941004@163.com})}

\thanks{Fuxin Ji, Dashan Jiang, Shuolin Liu,  and Qi Wu are with Department of Electrical Engineering and Automation, Anhui University, Hefei, China (e-mail: \mbox{2115560277@qq.com}; \mbox{384898642@qq.com}; \mbox{z18201021@stu.ahu.edu.cn};  \mbox{anhuiwuqi1994@163.com})}

\thanks{Hoyeon Lee is with Department of Radiation Oncology, Massachusetts General Hospital, Boston, MA, United States (e-mail: \mbox{leehoy12345@gmail.com})}

\thanks{Jun Lian is with Department of Radiation Oncology, University of North Carolina at Chapel Hill, Chapel Hill, NC, United States (e-mail: \mbox{jun\_lian@med.unc.edu})}

\thanks{Keng-Chi Liu is with Department of Medical Imaging, Taiwan AI Labs, Taipei, Taiwan (e-mail: \mbox{calvin89029@gmail.com})}

\thanks{Kentaro Miki is with Department Of Biomedical and Health Sciences , Hiroshima University, Hiroshima, Japan (e-mail: \mbox{kentaro-miki@hiroshima-u.ac.jp})}

\thanks{Dan Nguyen is with Medical Artificial Intelligence and Automation (MAIA) Laboratory, Department of Radiation Oncology, The University of Texas Southwestern Medical Center, Dallas, TX, United States (e-mail: \mbox{dan.nguyen@utsouthwestern.edu})}

\thanks{Hamidreza Nourzadeh is with Department of Radiation Oncology, Thomas Jefferson University, Philadelphia, PA, United States (e-mail: \mbox{hamidreza.nourzadeh@jefferson.edu})}

\thanks{Alexander F. I. Osman is with Department of Medical Physics, Al-Neelain University, Khartoum, Sudan (e-mail: \mbox{alexanderfadul@yahoo.com})}

\thanks{Hongming Shan is with Institute of Science and Technology for Brain-inspired Intelligence, Fudan University, Shanghai, China (e-mail: \mbox{hmshan@fudan.edu.cn})}

\thanks{Kay Sun is with Studio Vodels, Atlanta, GA, United States (e-mail: \mbox{nysuka@gmail.com})}

\thanks{Rob Verbeek is with Department Computer Science, Aalto University, Espoo, Finland (e-mail: \mbox{robzelluf@hotmail.com})}

\thanks{Siri Willems is with Department of Electrical Engineering, KULeuven, Leuven, Belgium (e-mail: \mbox{siri.willems@kuleuven.be})}

\thanks{Xuanang Xu is with Department of Biomedical Engineering, Rensselaer Polytechnic Institute, Troy, NY, United States (e-mail: \mbox{xux12@rpi.edu})}

\thanks{Sen Yang is with Tencent AI Lab, Shenzhen, Guangdong, China (e-mail: \mbox{ys810137152@gmail.com})}

\thanks{Lulin Yuan is with Department of Radiation Oncology, Virginia Commonwealth University Medical Center, Richmond, VA, United States (e-mail: \mbox{luliny@gmail.com})}

\thanks{Simeng Zhu is with Department of Radiation Oncology, Henry Ford Health System, Detroit, MI, United States (e-mail: \mbox{szhu1@hfhs.org})}

\thanks{Lukas Zimmermann is with Faculty of Health, University of Applied Sciences Wiener Neustadt, Wiener Neustadt, Austria and Competence Center for Preclinical Imaging and Biomedical Engineering, University of Applied Sciences Wiener Neustadt, Wiener Neustadt, Austria (e-mail: \mbox{Lukas.a.zimmermann@meduniwien.ac.at})}

\thanks{Kevin L. Moore is with Department of Radiation Oncology, University of California, San Diego, La Jolla, CA, United States (e-mail: \mbox{k3moore@health.ucsd.edu})}

\thanks{Thomas G.  Purdie is with Radiation Medicine Program, UHN Princess Margaret Cancer Centre, Toronto, ON, Canada, Department of Radiation Oncology, University of Toronto, Toronto, ON, Canada, Techna Institute for the Advancement of Technology for Health, Toronto, ON, Canada, and Department of Medical Biophysics, University of Toronto, Toronto, ON, Canada (e-mail: \mbox{Tom.Purdie@rmp.uhn.ca})}

\thanks{Andrea L. McNiven is with Radiation Medicine Program, UHN Princess Margaret Cancer Centre, Toronto, ON, Canada and Department of Radiation Oncology, University of Toronto, Toronto, ON, Canada (e-mail: \mbox{Andrea.McNiven@rmp.uhn.ca})}

\thanks{Timothy C. Y. Chan is with Department of Mechanical and Industrial Engineering, University of Toronto, Toronto, ON, Canada, Vector Institute, Toronto, ON, Canada, and Techna Institute for the Advancement of Technology for Health, Toronto, ON, Canada (e-mail: \mbox{tcychan@mie.utoronto.ca})}
}

	\bibliography{main}
	\bibliographystyle{IEEEtran}

\end{document}